\documentclass[prd,showpacs,showkeys,preprint,nofootinbib]{revtex4}
\usepackage{color}
\usepackage{amsmath}
\usepackage{amssymb}
\usepackage{yfonts}[1988/10/03]
\usepackage{graphicx}
\usepackage{subfig}
\newcommand {\beq}{\begin{equation}}
\newcommand {\eeq}{\end{equation}}
\newcommand {\beqa}{\begin{eqnarray}}
\newcommand {\eeqa}{\end{eqnarray}}

\begin{document}
	{\textsf{\today}
\title{ Gravitational Waves  with  Dark Energy }
\author{Jafar Khodagholizadeh }
\affiliation{ Farhangian University, P.O. Box 11876-13311,  Tehran,  Iran.
}

\begin{abstract}
In this article, we study the tensor mode equation of perturbation in the presence of nonzero$-\Lambda$ as dark energy, the dynamic nature of which depends on the Hubble parameter $ H$ and/or its time derivative. Dark energy, according to the total vacuum contribution, has a slight effect during the radiation-dominated era, but it reduces the squared amplitude of gravitational waves (GWs) up to $60\%$ for the wavelengths that enter the horizon during the matter-dominated era. Moreover, the observations bound on dark energy models, such as running vacuum model (RVM), generalized running vacuum model (GRVM) and generalized running vacuum subcase (GRVS), are effective in reducing the GWs’ amplitude. Although this effect is less for the wavelengths that enter the horizon at later times, this reduction is stable and permanent.   
\keywords{Gravitational waves, Dark energy}
\end{abstract}
\pacs{98.80.-k, 04.20.Cv, 02.40.-k}
\maketitle
\section*{Introduction}
Primordial gravitational waves (GWs; or tensor mode perturbations) arise from inflationary at early times and propagate freely to the expanding universe without any interactions with the matter and radiation \cite{Lifshitz, Mukhanov}. However, it is shown that at the temperature of $ \approx 10^{10}~(^{0}K)$, the tensor modes can be affected by anisotropic inertia, which contains free streaming neutrinos and antineutrinos \cite{damping}. This issue has been explored in the spatially curved spacetime and extended to the $ \Lambda$-dominated era \cite{Khodagholizadeh:2014ixa, Khodagholizadeh:2017ttk, khoda}. Here, we study the effect of dark energy on GWs.
\\
Current observational data indicate that the dark energy is a cosmological constant without evidence for its conclusiveness \cite{126,128,130,132,133,134,136,137,143,145,146}. Moreover, dark energy is one of the most compelling mysteries, which is thought to be behind the present cosmic acceleration \cite{Riess, Perl}.
The dynamic nature of $\Lambda$ could be conventionally achieved in three models: the first model has an explicit function of time, in which the most popular relation is in the form of the inverse power law as $\Lambda(t)\propto t^{-n}$ \cite{44,45,46,47,48}, and the other in this category is proposed as an exponential decay \cite{52}. The next $\Lambda-$ model could be expressed in terms of the scale factor $a$, in which the general form is $ \Lambda(a)=A a^{n}+B a^{m}$ or the relation consistent with the data is $ \Lambda(a)=A + B a^{-n}$. In spatially flat cosmology, $n\geq 1.6$ shows consistency with lensing data \cite{63}, while $n=2$ is the framework of the closed cosmology with the condition that matter/radiation densities are equal to the critical density at all times \cite{65}. The third class of expressions for $\Lambda$ is based on the Hubble parameter as the usual function of $ H^{n}$ and $ dH/dt$, in which $n=1$ is disqualified by cosmic microwave background (CMB) data \cite{74}. A general form of this model is known as the generalized running vacuum model (GRVM):
\begin{equation}\label{Hubble}
	\Lambda(H)=A+B H^{2}+ C\dot{H}
\end{equation}
Here, dot denotes the differentiation with respect to cosmic time. 
$B$ and $C$ are constants and dimensionless, and $A$ has the unit of length$^{-2}$. A classical running vacuum model (RVM) \cite{24, 25, 26, 27, 28} is a null value for $C$. Both RVM and GRVM appear to provide a better fit to the structure formation data \cite{sola1, sola2, sola3}. We shall refer to the latter as the generalized running vacuum subcase (GRVS) with $B=0$, which was investigated in ref.\cite{30} as a model with a variable dark energy equation-of-state parameter. Here, we assumed that the equation-of-state parameter of dark energy, $w$, is fixed at -1, similar to $\Lambda$CDM. In the following, we study the effect of the three models on GWs.
\\
This paper is organized as follows. In Sec. II, we derive the equation of the tensor mode perturbation in the presence of $\Lambda$ as dark energy. In the next Sections, the solutions of the mentioned equation are studied separately in all the cosmic history. Finally, the results are reported.
\section*{ Linear field equation with dynamic dark energy}
Let’s decompose the perturbed metric as:
\begin{eqnarray}
	g_{\mu\nu}=\bar{g}_{\mu\nu}+h_{\mu\nu},
\end{eqnarray}
where $\bar{g}_{\mu\nu}$ is the background metric and $h_{\mu\nu}$ is the symmetric perturbation term with the condition $\mid h_{\mu\nu}\mid\ll 1 $. The metric components of the Friedmann–Lemaitre–Robertson–Walker (FLRW) model in the Cartesian coordinate system are \cite{1}:
\begin{eqnarray}\label{metric}
	\bar{g}_{00}&=&-1,\;\;\;\;\bar{g}_{i0}=0,\;\;\;\bar{g}_{ij}=a^2(t)\tilde{g}_{ij}
	,\;\;\;\tilde{g}_{ij}=\delta_{ij}+K\frac{x^ix^j}{1-Kx^2},
\end{eqnarray}
with $i$ and $j $ running over the values 1, 2 and 3; $x^0=t $ is the time coordinate in our units, the speed of light is equal to unity and $K$ is the curvature constant. Also, $a(t)$ is the scale factor, which will be $\alpha\cosh(t/\alpha)$ and $ \alpha=\sqrt{\dfrac{3}{\Lambda}}$ in the closed de Sitter spacetime. From \cite{Pad}, the field equation for the tensor mode fluctuation in the source free region is:
\begin{eqnarray}\label{Eq1}
	\square h_{\mu\nu}+2 \bar{R}_{\mu\alpha \nu\beta}^{0} h^{\alpha\beta}=0 
\end{eqnarray}
where $\bar{R}_{\mu\alpha \nu\beta}^{0} $ is the background Riemann tensor. This equation describes the propagation of weak GWs in the source-free region of the curved spacetime. If the amplitude and the wavelength of $h^{\mu\nu}$ are $\epsilon$ and $ \lambda$, in the above relation, the first term is in the order of $ O(\epsilon/\lambda^{2}) $ and the second term is $O(\epsilon/ L^{2}) $ while  $L$ is the scale over which the background geometry is varied. Hence, in the order of $ O(\lambda^{2}/\epsilon^{2}) $, we can ignore the second term and the well-known GW equation will be: 
\begin{eqnarray}\label{13}
	\square h_{\mu\nu}=0 
\end{eqnarray}
For calculating, the metric can be put in the form of $ h_{ij}=a^2D_{ij}$ where $D_{ij}$s are the functions of $\vec{X}$ and $t$, satisfying the traceless-transverse conditions (or TT gauge):
\begin{eqnarray}
	\tilde{g}^{ij}D_{ij}=0,\;\;\; \tilde{g}^{ij}\bar{\nabla}_iD_{jk}=0.
\end{eqnarray}
These conditions are used to obtain Eq.(\ref{Eq1}). With manipulation calculations shown in Appendix A, we can show that:
\begin{eqnarray}
	\square h_{\mu\nu}&=&  \partial^{m}\partial_{m}D_{ij}-3 Kx^{m} \partial_{m}D_{ij}-K x^{m}x^{n}\partial^{m}\partial_{n}D_{ij}-2K x^{l}(\partial_{i}D_{lj}+\partial_{j}D_{il})\nonumber\\&+& 2K^{2}\tilde{g}_{ij}x^{k}x^{l}D_{kl} -a^{2}\ddot{D}_{ij}-3a\dot{a}\dot{D}_{ij}-2K D_{ij}+2\dot{a}^{2}D_{ij}
\end{eqnarray}
Moreover, one can show with straightforward calculations that: 
\begin{eqnarray}
	\nabla^2D_{jk}\equiv \bar{g}^{mn}\nabla_m\nabla_nD_{jk} &=&\partial^{m}\partial_{m}D_{ij}-3 Kx^{m} \partial_{m}D_{ij}-K x^{m}x^{n}\partial^{m}\partial_{n}D_{ij}\nonumber\\&-&2K x^{l}(\partial_{i}D_{lj}+\partial_{j}D_{il})+ 2K^{2}\tilde{g}_{ij}x^{k}x^{l}D_{kl} -2K D_{ij}
\end{eqnarray}
By using the above expression and the Friedmann equation, $ 2\dot{a}^2+a\ddot{a}=\Lambda(t) a^2-2K$, in which the cosmological constant is of dynamic nature, Eq.(\ref{13}) will be: 
\begin{eqnarray}
	\nabla^2D_{jk}-a^{2}\ddot{D}_{ij}-3a\dot{a}\dot{D}_{ij}+(\Lambda(t) a^{2}-a\ddot{a}-2K) D_{ij}=0
\end{eqnarray}
By looking at the plane-wave analogue and without losing generality, we chose the solution
of the above equation as a move in the $z-$ direction. Therefore, by using the modified transverse and traceless conditions, with some manipulation, we conclude that each mode, $+$ and $ \times $, holds true in the following relation   (e.g., in closed spacetime, $ K=1$; please see Relations (29) and (30) in Ref.\cite{Khodagholizadeh}).
\begin{eqnarray}
	&&(1-z^{2}) \dfrac{d^{2}}{dz^{2}}D(z,t)+3z \dfrac{d}{dz}D(z,t)-D(z,t)+\dfrac{6D(z,t)}{1-z^{2}} \nonumber\\
	&&-a^{2}\ddot{D}(z,t)- 3 a \dot{a} \dot{D}(z,t)+[\Lambda(t)a^{2}-a\ddot{a}-2K]D(z,t)=0 
\end{eqnarray}
With the method of separation of variables $D(z,t)=\hat{D}(z) D(t)$, the time evolution of the tensor mode perturbation will be:
\begin{eqnarray}
	\dfrac{a^{2}\ddot{D}(t)}{D(t)}+\dfrac{3a \dot{a} \dot{D}(t)}{D(t)}+ a\ddot{a}-\Lambda(t)a^{2}+2K=-m^{2}
\end{eqnarray}
where $m^{2}=n^{2}-3$ in which $n$ is a wavenumber. If the space-time background is curved, the wavenumber must be discrete, which comes from examining the periodic solution of the spatial part of the wave equation \cite{Khodagholizadeh}. Therefore, the final time evolution of GWs in the presence of $\Lambda(t)$ as dark energy will be: 
\begin{eqnarray}\label{final}
	\ddot{D}(t)+ 3 \dfrac{\dot{a}}{a}\dot{D}(t)+\dfrac{q^{2}}{a^{2}}D(t)=(\Lambda(t) - \dfrac{\ddot{a}}{a})D(t)
\end{eqnarray}
The expression $q^{2}= n^{2}-3+2K $ is also a wavenumber. It should be noted that, in Eq. (\ref{Eq1}), the presence of the second term results in the expression $ 2 a^{2}$, in which the value 2 is added to the discrete wavenumber $q^{2}$.
\\
Here, we have both the spatially curvature parameter and the cosmological constant. It seems the nonzero curvature has an effect on constraining some dark energy models \cite{Polarski, Franca, Ichikawa, Ichikawa1, Clarkson, Gong, Ichikawa2, Wright, Zhao}. Using CMB, type Supernova Ia (SNe Ia) and galaxy survey data show that the bounds on cosmic curvature are less stringent if dark energy density is allowed to be free of redshift and are dependent on the assumption about its early time properties. However, assuming a constant dark energy equation of state gives the most stringent constraints on cosmic curvature\cite{Wang:2007mza}. Nevertheless, for all the values of the curvature parameter, $\Lambda-$term has an important role in the evolution of the GWs.\\ 
In the next section, the treat of Eq.(\ref{final}) is studied in the radiation-dominated era.
\section*{
 Effect of  running vacuum model in radiation-dominated era} 
The  parameters $B$ and/or $ C $ are constrained by means of an Markov chain 
Monte Carlo (MCMC) analysis, initially using data for the observables associated with SNe Ia, cosmic chronometers, CMB and baryon acoustic oscillations (BAOs) \cite{Farrugia:2018mex}. At early times the Eq.(\ref{Hubble}) is no longer valid because the density of dark energy 'blows up' (it decays with time, so if you extend the model too far back, you get unrealistically large quantities). On the other hand, it is better to use a more general model.
The total vacuum contribution or the RVM with its generalized version takes spatial curvature into account after inflation is described in the complete cosmic history as \cite{Lima:2015kda}. 
\begin{eqnarray}
	\Lambda(H,a)=\Lambda_{\infty}+3\nu (H^{2}-H_{F}^{2}+\dfrac{K}{a^{2}})+3\alpha(\dfrac{H}{H_{I}})^{n}(H^{2}+\dfrac{K}{a^{2}})
\end{eqnarray}
where $H_{I} $ and $H_{F}$ stand for the Hubble parameter in two different epochs, while the former characterizes inflation, the latter denotes the final value of H as $a\longrightarrow \infty$. Also, $\Lambda_{\infty}$ is the limit of $\Lambda(H,a)$ as $ a \longrightarrow \infty$. Although rather generally, the above expression can be simplified based on different arguments. First, without loss of generality, we can see that parameter $ \alpha$ can be absorbed in the value of the scale $ H_{I}$, so that we may fix $\alpha=1$ and, with this condition, there is no fluid component; therefore, we have $ H=H_{I}$. Another reason is that $ H $ is expected to be already much smaller than $H_{I}$ at the beginning of adiabatic radiation phase and so the constant $\Lambda_{\infty}- $term dominates with the model following $\Lambda CDM$ evolution\cite{Lima:2015kda}. Since we are not concerned with inflation, but rather with the late time behavior of dark energy models, the term $(H/H_{I})^{n} $ may therefore be dropped and the resulting cosmology converge to $\Lambda CDM$. Moreover, in a flat case, S. Basilakos et. al \cite{Basil} obtained $ \nu\simeq 10^{-3}$ based on a joint analysis involving CMB, SNe Ia and BAO, while a theoretical analysis by J. Sola \cite{Sola} yielded $ |\nu|\sim 10^{-6}-10^{-3}$ within a generic grand unified theory (GUT). Therefore, since the curvature must be very small nowadays, $ \nu=0$ can be assumed for all values of the curvature. With the above explanations, assuming the simplest form in the radiation-dominated phase, in which dark energy is coupled to the matter, we have:    
\begin{eqnarray}
	\Lambda(t)=\Lambda_{\infty}+ 3 (H^{2}+\dfrac{K}{a^{2}})
\end{eqnarray} 
As mentioned,  $ \Lambda_{\infty}$  is the value of the running $\Lambda(H,a) $ when $ a\longrightarrow \infty$; thus, it can be neglected at early times; therefore, with $ \dfrac{\ddot{a}}{a}=\dot{H}+H^{2}$ and the above expression, Eq. (\ref{final}) will be:
\begin{eqnarray}\label{final1}
	\ddot{D}(t)+ 3 \dfrac{\dot{a}}{a}\dot{D}(t)+\dfrac{n^{2}-3-K}{a^{2}}D(t)=(2 H^{2}-\dot{H})D(t)
\end{eqnarray}
To investigate the treat of the tensor mode in the radiation- and matter-dominated eras, it is convenient to change the independent variable $t$ to $u= q\tau= q\int_{0}^{t} \dfrac{dt^{\prime}}{a(t^{\prime})}=\dfrac{2qt}{a(t)}$. By using the Friedmann equation $\dfrac{8\pi G \bar{\rho}}{3}=H^{2}=\dfrac{1}{4t^{2}}$, and $a(t)= t^{1/2}$ as the scale factor in the radiation-dominated era, Eq. (\ref{final1}) becomes
\begin{eqnarray}\label{24}
	\dfrac{d^{2}}{du^{2}} D_{n}(u)+ (\dfrac{2}{u}) \dfrac{d}{du} D_{n}(u)+D_{n}(u)=(\dfrac{2}{u^{2}}) D_{n}(u)
\end{eqnarray}
Generally, tensor mode perturbation rapidly becomes time-independent after the horizon exit and remains so until horizon re-entry; thus, there are initial conditions:
\begin{figure}
	\includegraphics[scale=0.7]{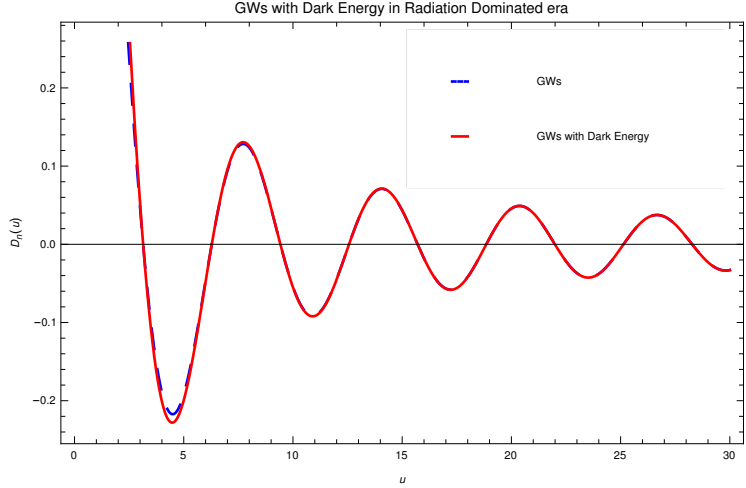}
	\caption{Comparing GWs in the presence or absence of dynamical dark energy (DE) in the radiation-dominated era shows the solid (blue) line is for GWs in the absence of DE and the dotted (orange) line is in the presence of DE. It seems that dark energy has no damping effect on GWs in the radiation epoch.}
	\label{Radiation}
\end{figure}
\begin{eqnarray}
	D_{n}(0)=1~~~~~~,~~~~~~\dfrac{d}{du}D_{n}(0)=0
\end{eqnarray}
For deep inside the horizon, Eq. (\ref{24}) approaches a solution as:
\begin{eqnarray}
	D_{n}(u)=D_{n}^0 j_1(x) + D_{n}^{1} y_1(x) 
\end{eqnarray}
where $ j_1(x) $ and $y_1(x) $ are the spherical Bessel functions of the first and second kinds, respectively. By using the definition of spherical Bessel functions, $ j_1(u)=\dfrac{\sin u}{u^2}-\dfrac{\cos u}{u}$ and $ y_1(u)=\dfrac{\sin u}{u}-\dfrac{\cos u}{u^2}$, the general solution is:
\begin{eqnarray}
	D_{n}(u)=[-D_{n}^1 + \dfrac{D_{n}^0}{u}]\dfrac{\sin u}{u}-[D_{n}^0+ \dfrac{D_{n}^1}{u}]\dfrac{\cos u}{u}
\end{eqnarray}
The coefficient of the second term must be zero; so, $D_{n}^0=-\dfrac{D_{n}^1}{u} $. In addition, for large $u (u\gg 1)$, the tensor modes are deep inside the horizon and the solution has to approach the homogeneous solution; thus, $D_{n}^{1}=-\dfrac{1}{2}$. Therefore, the final solution will be:   
\begin{eqnarray}
	D_{n}(u)= (1+ \dfrac{1}{u^2}) \dfrac{\sin u}{ u}
\end{eqnarray}
The second term, which is due to the presence of dark energy, is very smaller than the first term. As shown in Fig.\ref{Radiation}, it has a slight  effect on reducing the amplitude of GWs in the radiation-dominated era, so it can be ignored.
\section*{ Short wavelengths in matter-dominated era}
For the more detailed study of GWs in the presence of dark energy in the matter-dominant era, using the definition $u=q\tau=q\int_{0}^{t}\dfrac{dt^{'}}{a(t^{'})}=\dfrac{3qt}{a(t)}$, where $a(t)= t^{2/3}$, the equation of the tensor mode time evolution will be: 
\begin{eqnarray}\label{15}
	\dfrac{d^{2}}{du^{2}} D_{n}(u)+ (\dfrac{4}{u}) \dfrac{d}{du} D_{n}(u)+D_{n}(u)=(\dfrac{10}{u^{2}}) D_{n}(u)
\end{eqnarray}
The general solution is based on the Bessel functions of the first and second kinds as:     
\begin{eqnarray}
	D_n(u)=  \dfrac{1}{u^{3/2}}[D_{n}^{0}J_{\frac{7}{2}}(u)+ D_{n}^1 Y_{\frac{7}{2}}(u)]
\end{eqnarray}
where $D_{n}^{0}$ and $D_{n}^{0}$ are constant. At the moment that the perturbations enter the horizon, $u\cong 1$, the solution tends toward the solution of homogeneous equation; thus, it will be:
\begin{eqnarray}
	D_{n}(u)=\left(D_{n}^{0}-2Ci(2u)+\frac{0.48}{u^2}\right)\dfrac{\sin u}{u^2}
\end{eqnarray}
\begin{figure}
	\includegraphics[scale=0.8]{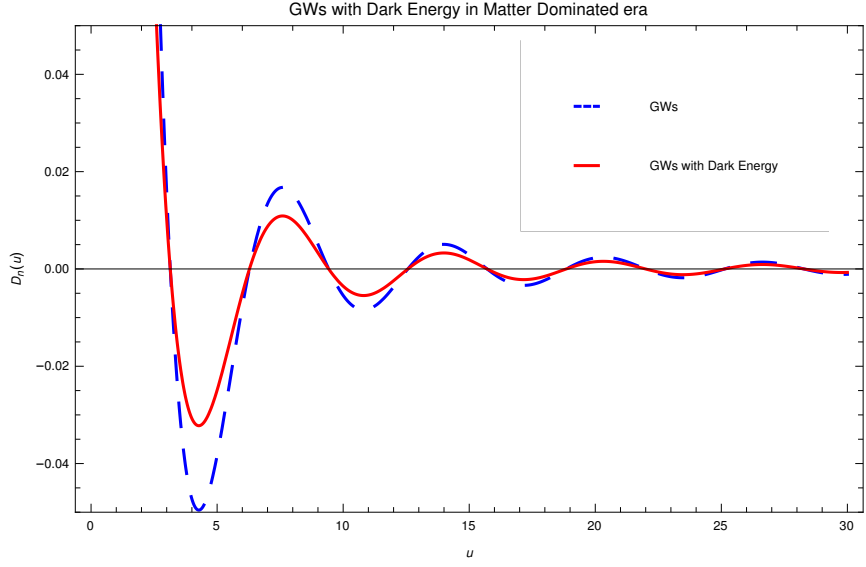}
	\caption{Comparing GWs in the presence or absence of dynamical dark energy in the matter-dominated era shows the dashed (blue) part is for GWs in the absence of DE and the dotted (orange) line is in the presence of DE.}
	\label{Matter1}
\end{figure}
where $ Ci(2u)$ is the cosine integral as $ Ci(2u)=\gamma +\ln(2u)+ \int_{0}^{2u}\dfrac{\cos t -1}{t} dt$. Deep inside the horizon, when $u\gg 1$, the right hand side of Eq. (\ref{15}) becomes negligible and the solution approaches a homogeneous solution as $\dfrac{\sin u}{u^2}$. Because large $u$, $ Ci(2u)$ and the third term of the general solution tend toward zero, $D_{n}^{0}=1$. As compared with the solution $\dfrac{\sin u}{u^2}$ in the absence of dark energy, Eq. (\ref{15}) shows that $ D_{n}(u)$ follows the without dark energy solution rather accurately until $ u\approx 1$, when the perturbation enters the horizon and, thereafter, rapidly approaches $ \approx 0.63 \dfrac{\sin(u+\delta)}{u^2}$, in which $ \delta$ is very small and negligible. Furthermore, it has a significant effect on decreasing the $ \textquoteleft\textquoteleft$B-B$\textquotedblright$ polarization multipole coefficient, $ C_{lB}$, which is up to $60\%$ less than it would be without the damping due to dark energy (see Fig. \ref{Matter1}).\\
As mentioned earlier, the constant parameters of RVM, GRVM and GRVS are obtained using the combination of type-Ia supernova, cosmic chronometers, BAO and CMB \cite{Farrugia:2018mex}. The results are based on the data sets which are in the framework of $\Lambda CDM$ cosmology with freely varying $\Omega_{K}^{0}$ and are valid Until near the  end of matter-dominated era. A general form of the GW equation with the mentioned models will be:
\begin{eqnarray}\label{Final2}
	\dfrac{d^{2}}{du^{2}} D_{n}(u)+ (\dfrac{4}{u}) \dfrac{d}{du} D_{n}(u)+D_{n}(u)=-(\dfrac{2-4 B+6C}{u^{2}}) D_{n}(u)
\end{eqnarray}
The largest constant obtained based on models for RVM is $10^{3}B=4.05 _{-0.2043}^{+0.0052}$; for GRVM, it is $B=0.3590_{-0.2050}^{+0.6489}$ and  $C=0.228_{-0.1312}^{+0.4148}$; for GRVS, the biggest coefficient is $10^{3}C=4.9508_{-2.3946}^{+2.6763}$\cite{Farrugia:2018mex}.\\
For GRVM, RVM and GRVS, the numerical solutions of (\ref{Final2}), which are deep inside the horizon, are $ 0.76 \dfrac{\sin u}{u^2}$, $ 0.83 \dfrac{\sin u}{u^2}$ and $ 0.85 \dfrac{\sin u}{u^2}$, respectively. Thus, the dark energy decreases the $ \textquoteleft\textquoteleft$B-B$\textquotedblright$ polarization multipole coefficient, $ C_{lB}$, up to $42\%$, $31\%$ and $27\%$, respectively, less than it would be without the damping due to dark energy. Therefore, the effect of dark energy on reducing the amplitude will be less when the expression $2-4 B+6C$ is larger.
In other words, the effect of dark energy on cosmological GWs will be greater if, in all theories, the value of $ B$ is higher or the value of $C$ is lower.
\section*{ General wavelengths in the dark energy-dominated era} 
After $ Z \lesssim 0.5$, the universe begins to accelerate and the dark energy is dominated \cite{Frieman:2008sn}. To investigate tensor mode perturbation, it is convenient to change the independent variable $ t$ to $ \chi =\dfrac{\bar{\rho}_{\Lambda}}{\bar{\rho}_{M}}= \dfrac{\bar{\rho}_{\Lambda, EQ}}{\bar{\rho}_{M,EQ}}\dfrac{a^{3}}{a_{EQ}^{3}}$
where $a_{EQ}$, $\bar{\rho}_{\Lambda, EQ}$ and $\bar{\rho}_{M, EQ} $ are the value of the Robertson-Walker scale factor, energy densities of vacuum and matter at matter-vacuum equality. Moreover, we have $ \dfrac{\Omega_{\Lambda}}{\Omega_{M}}=(1+Z_{EQ})^{3}$. $H_{EQ}=H_{0}\sqrt{2\Omega_{M}}(1+Z_{EQ})^{3/2}$ and $Z_{EQ}$ are the Hubble rate and the redshift at matter-vacuum equality, respectively.\\
In the condition $\chi \gg 1 $, when the dark energy is important, according to the Friedmann equation, we have:
\begin{eqnarray}
	H_{EQ} \dfrac{dt}{\sqrt{2}}=\dfrac{d\chi}{\chi\sqrt{1-\Omega_{K,EQ}}}
\end{eqnarray} 
where $\Omega_{K, EQ}$ is the curvature density at matter-vacuum equality. Thus, the homogeneous equation of GWs in the expansion universe will be:
\begin{eqnarray}\label{30}
	\dfrac{d^{2}}{d\chi^{2}}D_{n}(\chi)+ \dfrac{2}{\chi} \dfrac{d}{d\chi} D_{n}(\chi)+\dfrac{\kappa^{'^{2}}}{\chi^{2/3}}D_{n}(\chi)=0
\end{eqnarray} 
where $ \kappa^{'}$ is the dimensionless rescaled wavenumber:
\begin{eqnarray}
	\kappa^{'^{2}} =\dfrac{2q^{2}}{9 H_{EQ}^{2} a_{EQ}^{2}(1-\Omega_{K,EQ})}.
\end{eqnarray}
Due to the very small value of $\Omega_{K,EQ}$, e.g., $\Omega_{K,EQ}=(1+0.5)^{2}\Omega_{K}^{0}$ with $\Omega_{K}^{0}=-0.0001_{-0.0052}^{+0.0054}$ \cite{Planck:2015fie}, we can ignore it. Therefore:
\begin{eqnarray}
	\kappa^{'}=\dfrac{6.43(q/a_{0})}{\Omega_{M}h^{2}}[Mpc^{-1}]
\end{eqnarray} 
where $a_{0}$ is the present-day scale factor.
In the cases of long and short wavelengths, we have $\kappa^{'}\ll 1$ and $\kappa^{'}\gg 1$, respectively, and the cosmological GWs are detectable when $\kappa^{'}\gg 1$.\\
Recently, the existing ground-based operators, LIGO and VIRGO, reported GWs that are coming from black-hole mergers \cite{GWs}. These detectors do not have sensitivity to detect cosmological GWs; thus, they might be detected by space-borne laser interferometers, which operate at frequencies around $0.01$ to $0.1$ Hz. Therefore, GWs with the observed frequency of $qc/2\pi a_{0}=10^{-2}$ Hz \cite{Komatso} would have $\kappa^{'}\cong 0.43 \times 10^{15}/\Omega_{M}h^{2} \gg 1$ at the present epoch.\\ The damping effect is small in any way for $ \kappa^{'}\ll 1$; so, it will be adequate approximation for all the wavelengths to take the solution of Eq. (\ref{30}) in the $\Lambda$-dominated era to be given by multiplying  by a factor $\xi(\kappa^{'}) $:
\begin{eqnarray}
	D_{n}(\chi)=\xi(\kappa^{'}) \dfrac{3 \kappa^{'}}{\chi^{1/3}} \sin(\dfrac{3 \kappa^{'}}{\chi^{1/3}})
\end{eqnarray} 
where $\xi(\kappa^{'}) =\dfrac{1+ 0.76\kappa^{'}}{1+\kappa^{'}}$ in GRVM which is the biggest reduction. For $ \kappa^{'}\ll 1$, we have $\xi(\kappa^{'})=1$ and in $\kappa^{'}\gg 1$, the above relation will be:
\begin{eqnarray}
	D_{n}(\chi)=0.76\dfrac{3 \kappa^{'}}{\chi^{1/3}} \sin(\dfrac{3 \kappa^{'}}{\chi^{1/3}})
\end{eqnarray} 
\begin{figure} 
	{
		\includegraphics[width=0.47\textwidth]{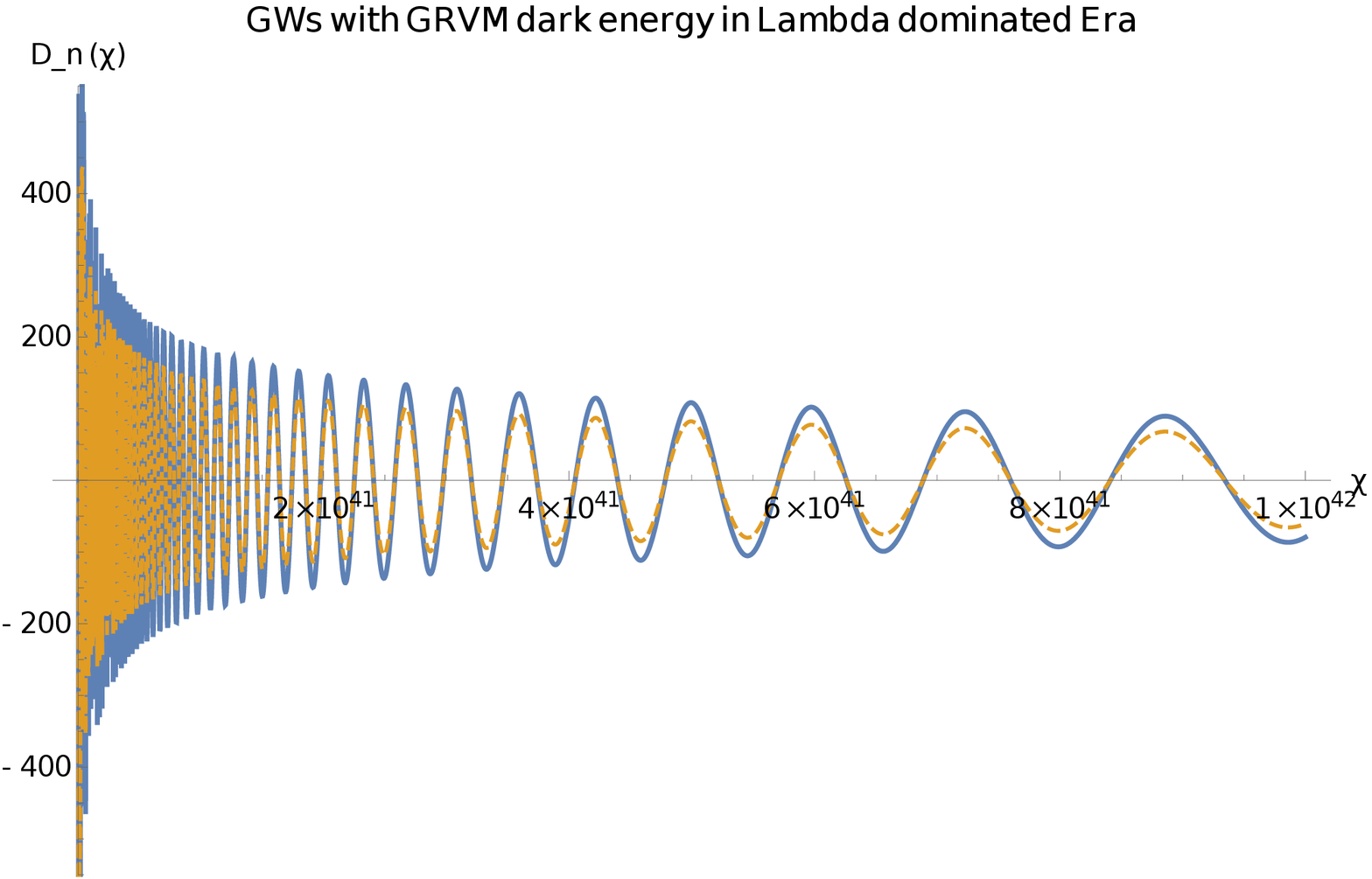} 
	} 
	\hfill 
	{ 
		\includegraphics[width=0.47\textwidth]{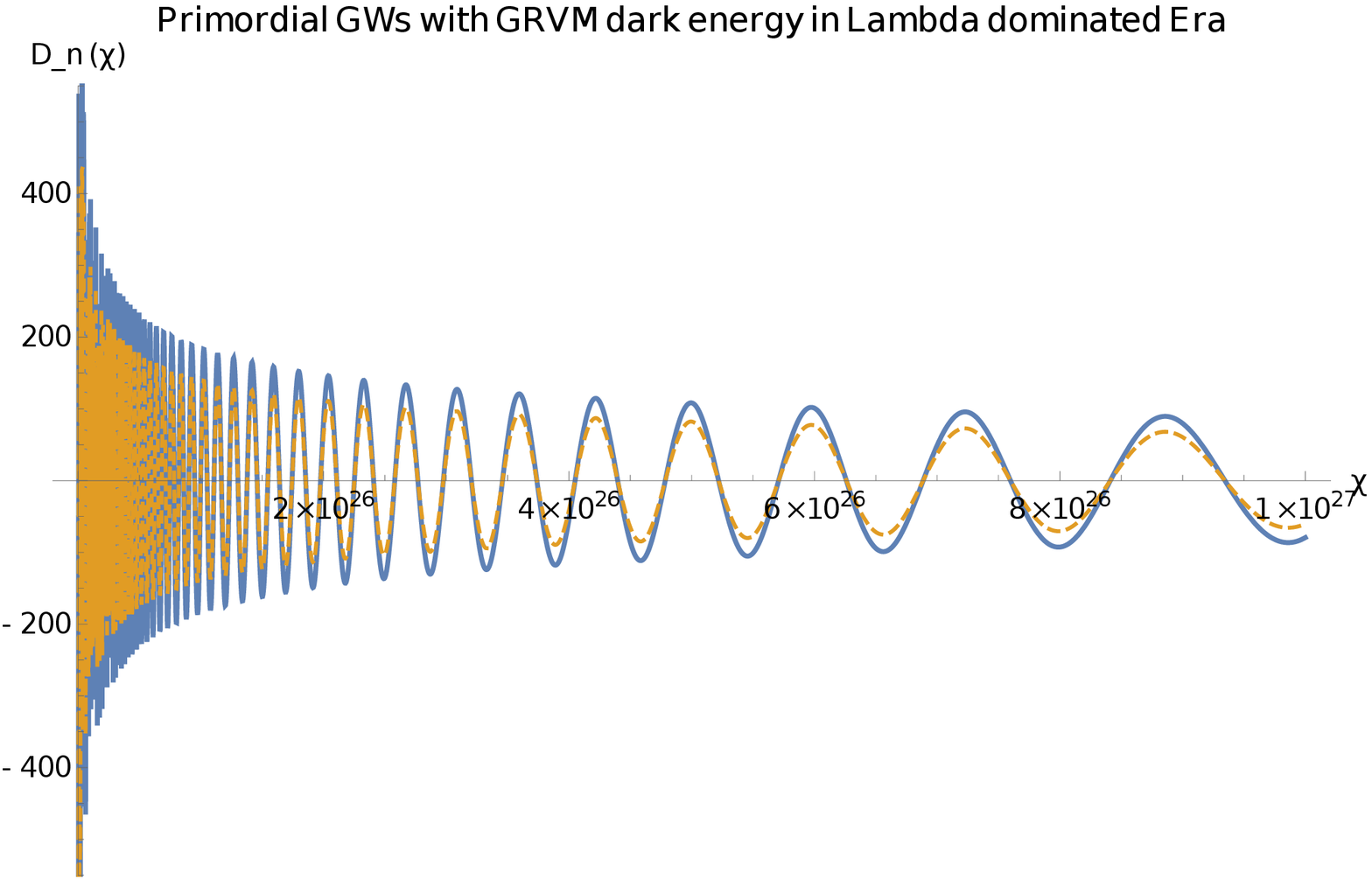} 
	} 
	\caption{Evolution of cosmological gravitational waves, $ D_n(\chi) $, with respect to the $\Lambda$-matter equality parameter $\chi$ is plotted in $ \Omega_{M}h^{2}=0.15$.  Left: In the presence of GRVM dark energy (dashed), the amplitude of GWs is less than the case without them (solid or blue line) at the observed frequency $\sim 10^{-2}$. Right: Stability is almost seen in the difference between the two states at the observed frequency $\sim 10^{-7}$ of the primordial gravitational waves coming from the quark-qluon plasma phase transition.} 
	\label{L}
\end{figure}
All the observable effects of primordial gravitational waves will be reduced by these factors of dark energy models. When a GW enters the horizon, it has short wavelength, but deep inside the horizon, it will take similar long wavelength (see Fig. \ref{L} Left).\\
If the frequency sensitivity of the detector is of $ 10^{-7} $ Hz which is able to observe the primordial gravitational waves coming from the quark-gluon plasma phase transition, then, $\kappa^{'}\cong 0.43 \times 10^{10}/\Omega_{M}h^{2} $. However, in the dark energy-dominated era, unlike the previous eras, such as the radiation and matter eras, this amplitude reduction which is due to dark energy relatively affects the long wavelengths (see the right plot in Fig. \ref{L}  ).
\section*{Conclusion}
In all scales and at all times, dark energy is present alongside GWs after the inflation epoch. Its minimum effect is seen in the radiation-dominated era, in which its solution will be in terms of $\dfrac{1}{u^{2}}$ multiplied by the spherical Bessel functions. In the matter-dominated era, according to the total vacuum contribution, we will have the maximum GW reduction as $0.63 \dfrac{\sin u}{u^2}$, but in the near end of its epoch, by using dark energy models such as RVM, GRVM and GRVS, the highest reduction is $0.76 \dfrac{\sin u}{u^2}$ and selecting the values of $B$ and $C$ is effective in the range of reduction. Hence, all the quadratic effects of the tensor modes in the CMB, such as tensor contribution to the temperature multipole coefficients $C_{l}$ and all of the $ \textquoteleft\textquoteleft$B-B$\textquotedblright$ polarization multipole coefficients, are $60\%$ less than they would be in the case without the damping due to dark energy terms with total vacuum contribution. The maximum reduction for polarization multipole coefficients is observed in the GRVM, which is $42\%$.  \\
In the dark energy-dominated era, unlike the epoch of the radiation- and matter-dominated eras, in which the effect of dark energy on GWs is destroyed at long wavelengths, this effect always exists and the amplitude reduction is stable. Also, the accommodation of nonzero spatial curvature with dark energy has almost no effect on GWs.

\section*{Appendix A} 
For calculating the $ \square h_{\mu\nu}$ we have
\begin{eqnarray}
	\square h_{\mu\nu}&=&\bigtriangledown_{\alpha}g^{\alpha\beta}\bigtriangledown_{\beta}h_{\mu\nu}=\bigtriangledown_{\alpha}g^{\alpha\beta}(\partial_{\beta}h_{\mu\nu}-\Gamma_{\beta\mu}^{\xi}h_{\xi\nu}-\Gamma_{\beta\nu}^{\xi}h_{\xi\mu})\nonumber\\&=& \partial_{\alpha}[g^{\alpha\beta}(\partial_{\beta}h_{\mu\nu}-\Gamma_{\beta\mu}^{\xi}h_{\xi\nu}-\Gamma_{\beta\nu}^{\xi}h_{\xi\mu})]+\Gamma_{\alpha\lambda}^{\alpha}g^{\lambda\beta}(\partial_{\beta}h_{\mu\nu}-\Gamma_{\beta\mu}^{\xi}h_{\xi\nu}-\Gamma_{\beta\nu}^{\xi}h_{\xi\mu})\nonumber\\&-& \Gamma_{\alpha\mu}^{\kappa}g^{\alpha\beta}(\partial_{\beta}h_{\kappa\nu}-\Gamma_{\beta\kappa}^{\xi}h_{\xi\nu}-\Gamma_{\beta\nu}^{\xi}h_{\xi\kappa})-\Gamma_{\alpha\nu}^{\kappa}g^{\alpha\beta}(\partial_{\beta}h_{\kappa\mu}-\Gamma_{\beta\mu}^{\xi}h_{\xi\kappa}-\Gamma_{\beta\kappa}^{\xi}h_{\xi\mu})
\end{eqnarray}
From the background metric (\ref{metric}) in addition to the perturbed metric, $ h_{ij}$,  the first term  will be :
\begin{eqnarray}\label{1}
	\partial_{\alpha}[g^{\alpha\beta}(\partial_{\beta}h_{\mu\nu}-\Gamma_{\beta\mu}^{\xi}h_{\xi\nu}-\Gamma_{\beta\nu}^{\xi}h_{\xi\mu})]&=&-\partial_{0}(\partial_{0}h_{\mu\nu}-\Gamma_{0\mu}^{\xi}h_{\xi\nu}-\Gamma_{0\nu}^{\xi}h_{\xi\mu})\nonumber\\&+&\partial_{m}g^{mn}(\partial_{n}h_{\mu\nu}-\Gamma_{n\mu}^{\xi}h_{\xi\nu}-\Gamma_{n\nu}^{\xi}h_{\xi\mu})\nonumber\\&=&-\partial_{0}(\partial_{0}h_{ij}-\dfrac{\dot{a}}{a}\delta_{il}h_{lj}-\dfrac{\dot{a}}{a}h_{li})\nonumber\\&&~+a^{-2}\{\partial_{m}(\tilde{g}^{mn}\partial_{n}h_{ij}-Kh_{ij}-Kx^{l}\partial_{i}h_{li} -K h_{ij}-Kx^{l}\partial_{i}h_{li})\}.\nonumber\\
\end{eqnarray}
The second term will be 
\begin{eqnarray}\label{2}
	\Gamma_{\alpha\lambda}^{\alpha}g^{\lambda\beta}(\partial_{\beta}h_{\mu\nu}-\Gamma_{\beta\mu}^{\xi}h_{\xi\nu}-\Gamma_{\beta\nu}^{\xi}h_{\xi\mu})&=&\Gamma_{mn}^{m}g^{nk}(\partial_{k}h_{\mu\nu}\nonumber\\&-&\Gamma_{k\mu}^{\xi}h_{\xi\nu}-\Gamma_{k\nu}^{\xi}h_{\xi\mu})-\Gamma_{m0}^{m}(\partial_{0}h_{\mu\nu}-\Gamma_{0\mu}^{\xi}h_{\xi\nu}-\Gamma_{0\nu}^{\xi}h_{\xi\mu})\nonumber\\&=& Kx^{m} a^{-2}(\partial_{m}h_{ij}-K\tilde{g}_{mi}x^{m}x^{l}h_{lj}-K^{2}\tilde{g}_{mj}x^{m}x^{l}h_{il})\nonumber\\&-&3\dfrac{\dot{a}}{a}(\partial_{0}h_{ij}-2\dfrac{\dot{a}}{a}h_{ij})\nonumber\\
\end{eqnarray}
The third term will be 
\begin{eqnarray}\label{3th}
	-\Gamma_{\alpha\mu}^{\kappa}g^{\alpha\beta}(\partial_{\beta}h_{\kappa\nu}-\Gamma_{\beta\kappa}^{\xi}h_{\xi\nu}-\Gamma_{\beta\nu}^{\xi}h_{\xi\kappa})&=& \Gamma_{0\mu}^{\kappa}(\partial_{0}h_{\kappa\nu}-\Gamma_{0\kappa}^{\xi}h_{\xi\nu}-\Gamma_{0\nu}^{\xi}h_{\xi\kappa})\nonumber\\&-& \Gamma_{m\mu}^{\kappa}g^{mn}(\partial_{n}h_{\kappa\nu}-\Gamma_{n\kappa}^{\xi}h_{\xi\nu}-\Gamma_{n\nu}^{\xi}h_{\xi\kappa})\nonumber\\&=&  \dfrac{\dot{a}}{a}(\partial_{0}h_{ij}-2\dfrac{\dot{a}}{a}h_{ij})-a \dot{a}\tilde{g}_{mi}a^{-2} \tilde{g}^{mn}\Gamma_{n0}^{l}h_{lj} \nonumber\\&&~~~~~~~- Kx^{k} a^{-2}(\partial_{i}h_{jk}-Kx^{l}\tilde{g}_{ik}h_{lj}-Kx^{l}\tilde{g}_{ij}h_{kl})
\end{eqnarray}
and the  fourth term will be 
\begin{eqnarray}\label{4th}
	-\Gamma_{\alpha\nu}^{\kappa}g^{\alpha\beta}(\partial_{\beta}h_{\kappa\mu}-\Gamma_{\beta\mu}^{\xi}h_{\xi\kappa}-\Gamma_{\beta\kappa}^{\xi}h_{\xi\mu})&=&\Gamma_{0\nu}^{\kappa}(\partial_{0}h_{\kappa\mu}-\Gamma_{0\mu}^{\xi}h_{\xi\kappa}-\Gamma_{0\kappa}^{\xi}h_{\xi\mu})\nonumber\\&-&\Gamma_{m\nu}^{\kappa}g^{mn}(\partial_{n}h_{\kappa\mu}-\Gamma_{n\mu}^{\xi}h_{\xi\kappa}-\Gamma_{n\kappa}^{\xi}h_{\xi\mu})\nonumber\\&=& \dfrac{\dot{a}}{a}(\partial_{0}h_{ij}-2\dfrac{\dot{a}}{a}h_{ij})+\dfrac{\dot{a}}{a} \tilde{g}_{mj}\tilde{g}^{mn}\Gamma_{n0}^{l}h_{il} \nonumber\\&&~~~~~~~- Kx^{k} a^{-2}(\partial_{j}h_{ik}-Kx^{l}\tilde{g}_{ij}h_{lk}-Kx^{l}\tilde{g}_{kj}h_{il})\nonumber\\
\end{eqnarray}
The tensor mode perturbation to the metric can be put in the form
$ h_{ij}=a^2D_{ij}$ which $h_{00}=0 $ and $ h_{ij}=0$, 
where $D_{ij}$s are functions of $\vec{X}$ and $t$, satisfying the traceless-transverse conditions (or  TT gauge):
\begin{eqnarray}
	\tilde{g}^{ij}D_{ij}=0,\;\;\; \tilde{g}^{ij}\bar{\nabla}_iD_{jk}=0.
\end{eqnarray}
Therefore with manipulation calculations by the  summation of relations (\ref{1}), (\ref{2}),(\ref{3th}) and (\ref{4th}) we have
\begin{eqnarray}
	\square  h_{ij}&=&-2a\dot{a}\dot{D}_{ij}-a^{2}\ddot{D}_{ij}+\partial^{m}\partial_{m}D_{ij}-4K x^{m}\partial_{m}D_{ij}-K x^{m}x^{n}\partial_{n}\partial_{m}D_{ij}-2K D_{ij}\nonumber\\&-&K x^{l}(\partial_{i}D_{lj}+\partial_{j}D_{il})+ K x^{m}\partial_{m}D_{ij}-K^{2}\tilde{g}_{mi}x^{m}x^{l}D_{lj}-K^{2}\tilde{g}_{mij}x^{m}x^{l}D_{il}-3a \dot{a}\dot{D}_{ij}\nonumber\\&+& \dot{a}^{2}D_{ij}-Kx^{m}\partial_{i}D_{jm}+K^{2}\tilde{g}_{im}x^{m}x^{l}D_{lj}+K^{2}\tilde{g}_{ij}x^{m}x^{l}D_{ml}+a\dot{a} \dot{D}_{ij}+\dot{a}^{2}D_{ij}\nonumber\\&-&Kx^{m}\partial_{j}D_{im}+K^{2} x^{m}x^{l}\tilde{g}_{ij} D_{ml} +K^{2} x^{m}x^{l}\tilde{g}_{jm} D_{il}
\end{eqnarray}
Then
\begin{eqnarray}
	\square h_{\mu\nu}&=&  \partial^{m}\partial_{m}D_{ij}-3 Kx^{m} \partial_{m}D_{ij}-K x^{m}x^{n}\partial^{m}\partial_{n}D_{ij}-2K x^{l}(\partial_{i}D_{lj}+\partial_{j}D_{il})\nonumber\\&+& 2K^{2}\tilde{g}_{ij}x^{k}x^{l}D_{kl} -a^{2}\ddot{D}_{ij}-3a\dot{a}\dot{D}_{ij}-2K D_{ij}+2\dot{a}^{2}D_{ij}
\end{eqnarray}
Also one can show with straightforward calculations 
\begin{eqnarray}
	\nabla^2D_{jk}\equiv \bar{g}^{mn}\nabla_m\nabla_nD_{jk} &=&\partial^{m}\partial_{m}D_{ij}-3 Kx^{m} \partial_{m}D_{ij}-K x^{m}x^{n}\partial^{m}\partial_{n}D_{ij}\nonumber\\&-&2K x^{l}(\partial_{i}D_{lj}+\partial_{j}D_{il})+ 2K^{2}\tilde{g}_{ij}x^{k}x^{l}D_{kl} -2K D_{ij}
\end{eqnarray}
By using above expression and  the Friedmann equation, $ 2\dot{a}^2+a\ddot{a}=\Lambda(t) a^2-2K$ in which cosmological constant has a dynamic nature,   the Eq.(\ref{13}) will be 
\begin{eqnarray}
	\nabla^2D_{jk}-a^{2}\ddot{D}_{ij}-3a\dot{a}\dot{D}_{ij}+[ \Lambda(t) a^{2}-a\ddot{a}-2K] D_{ij}=0
\end{eqnarray}

\end{document}